\documentclass{cernrep}

\usepackage{fancyhdr}
\fancyhfoffset{4 mm}
\fancypagestyle{ARTTITLE}{%
\fancyhf{} 
\lhead{\small{Proceedings of the CERN--Accelerator--School course: \it{Introduction to Accelerator Physics}}}
\lfoot{Available online at \url{https://cas.web.cern.ch/previous-schools}}
\rfoot{\thepage\hspace*{3mm}}

}

\usepackage[colorlinks=true, linkcolor=blue, citecolor=black, filecolor=black, urlcolor=blue]{hyperref}

\usepackage{epsfig}
\usepackage{listings}
\usepackage{hyperref}
\usepackage{tcolorbox}
\usepackage{lipsum}
\usepackage{amsmath}

\definecolor{codegreen}{rgb}{0,0.6,0}
\definecolor{codegray}{rgb}{0.5,0.5,0.5}
\definecolor{codepurple}{rgb}{0.58,0,0.82}
\definecolor{backcolour}{rgb}{0.95,0.95,0.92}

\lstdefinestyle{mystyle}{
    backgroundcolor=\color{backcolour},   
    commentstyle=\color{codegreen},
    keywordstyle=\color{magenta},
    numberstyle=\tiny\color{codegray},
    stringstyle=\color{codepurple},
    basicstyle=\footnotesize,
    breakatwhitespace=false,         
    breaklines=true,                 
    captionpos=b,                    
    keepspaces=true,                 
    numbers=left,                    
    numbersep=5pt,                  
    showspaces=false,                
    showstringspaces=false,
    showtabs=false,                  
    tabsize=2,
    basicstyle=\ttfamily\footnotesize
}

    {\endtcolorbox}

    {\endtcolorbox}

    {\endtcolorbox} 
 
\lstdefinelanguage{Maxima}{
keywords={addrow,addcol,zeromatrix,ident,augcoefmatrix,ratsubst,diff,ev,tex,%
with_stdout,nouns,express,depends,load,submatrix,div,grad,curl,%
rootscontract,solve,part,assume,sqrt,integrate,abs,inf,exp},
sensitive=true,
comment=[n][\itshape]{/*}{*/}
}

\begin{document}
\title{Tools for Scientific Computing}
\author{A. Latina, CERN}
\begin{abstract}
A large multitude of scientific computing tools is available today. This article gives an overview of available tools and explains the main application fields. In addition basic principles of number representations in computing and the resulting truncation errors are treated.
The selection of tools is for those students, who work in the field of accelerator beam dynamics.
\end{abstract}
\maketitle
\section{Introduction}
\maketitle
\thispagestyle{ARTTITLE}

In this lesson, we will outline good practice in scientific computing and guide the novice through the multitude of tools available. We will describe the main tools and explain which tool should be used for a specific purpose, dispelling common misconceptions.

We will suggest reference readings and clarify important aspects of numerical stability to help avoid making bad but unfortunately common mistakes. Numerical stability is the basic knowledge of every computational scientist.

We will exclusively refer to free and open-source software running on Linux or other Unix-like operating systems. Also, we will unveil powerful shell commands that can speed up simulations, facilitate data processing, and in short, increase your scientific throughput. 

\subsection{Floating-point numbers}
Computers store numbers not with infinite precision but rather in some approximation that can be packed into a fixed number of bits. One commonly encounters \emph{integers} and \emph{floating-point} numbers.

The technical standard for floating-point arithmetic,
IEEE Standard for Floating-Point Arithmetic (IEEE-754),
was established in 1985 by the Institute of Electrical and Electronics Engineers (IEEE). The standard addressed many problems found in the diverse floating-point implementations that made them difficult to use reliably and portably. 

\subsubsection{Error, accuracy, stability}
Arithmetic between numbers in {integer} representation is \emph{exact}, if the answer is not outside the range of integers that can be represented. \emph{Real} numbers use a floating-point representation IEEE-754, where a number is represented internally in scientific binary format, by a sign bit (interpreted as plus or minus), an exact integer exponent, and an exact positive integer mantissa (or fraction), such that a number is written as:
$$\textnormal{value} = \left(-1\right)^\textnormal{sign}\times 2^\textnormal{exponent}\times1.\textnormal{M}$$ 
\begin{figure}[!htb]
    \centering
    \includegraphics[width=0.9\textwidth]{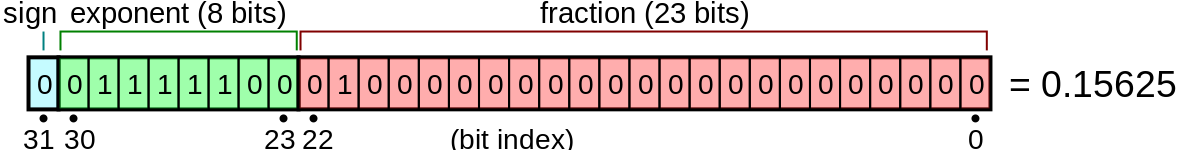}
    \caption{Single-precision floating point representation (32 bits).}
\end{figure}
\begin{figure}[!htb]
    \centering
    \includegraphics[width=0.9\textwidth]{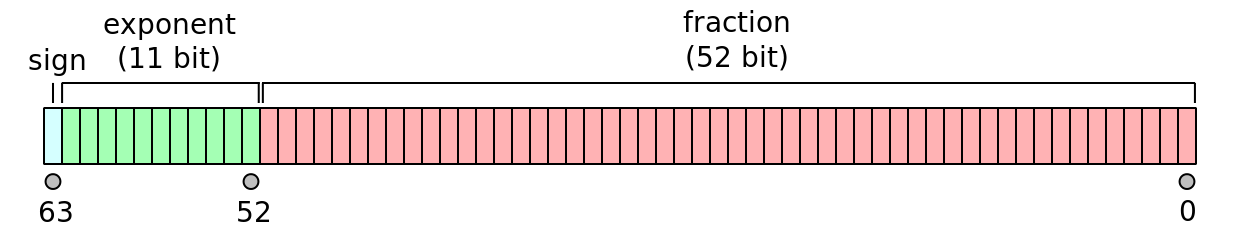}
    \caption{Double-precision floating point representation (64 bits).}
\end{figure}
\emph{Single-precision} floating point numbers use 8 bits for the exponent (therefore one can have exponents between -128 and 127) and 23 bits for the mantissa, $M$. \emph{Double-precision} floating point numbers use 11 bits for the exponent (one can have exponents between -1024 and 1023) and 52 bits for the mantissa, $M$.

The complete range of the positive normal floating-point numbers in \emph{single-precision} format is:
\begin{align*}
    s_\text{min} &= 2^{-126} \approx 1.17\times10^{-38},\\
    s_\text{max} &= 2^{127} \approx 3.4\times10^{38}.\\
\end{align*}
In \emph{double-precision} format the range is:
\begin{align*}
    d_\text{min} &= 2^{-1022} \approx  2\times10^{-308},\\
    d_\text{max} &= 2^{1024} \approx 2\times10^{308}.\\
\end{align*}
Some CPUs internally store floating point numbers in even higher precision: 80-bit in \emph{extended} precision, and 128-bit in \emph{quadrupole} precision. In C++ quadrupole precision may be specified using the \verb|long double| type, but this is not required by the language (which only requires \verb|long double| to be at least as precise as double), nor is it common. For instance, \verb|gcc| implements \verb|long double| as extended precision.

\subsubsection{Machine accuracy and round-off error}
The smallest floating-point number which, when added to the floating-point number 1.0, produces a floating-point result different from 1.0 is termed the \emph{machine accuracy} $\varepsilon_m$. For single precision $$\varepsilon_m \approx 3\cdot10^{-8},$$for double precision $$\varepsilon_m \approx 2\cdot10^{-16}.$$

It is important to understand that $\varepsilon_m$ is not the smallest floating-point number that can be represented on a machine. That number depends on how many bits there are in the exponent, while $\varepsilon_m$ depends on how many bits there are in the mantissa.

\begin{figure}[!htb]
    \centering
    \includegraphics[width=\textwidth]{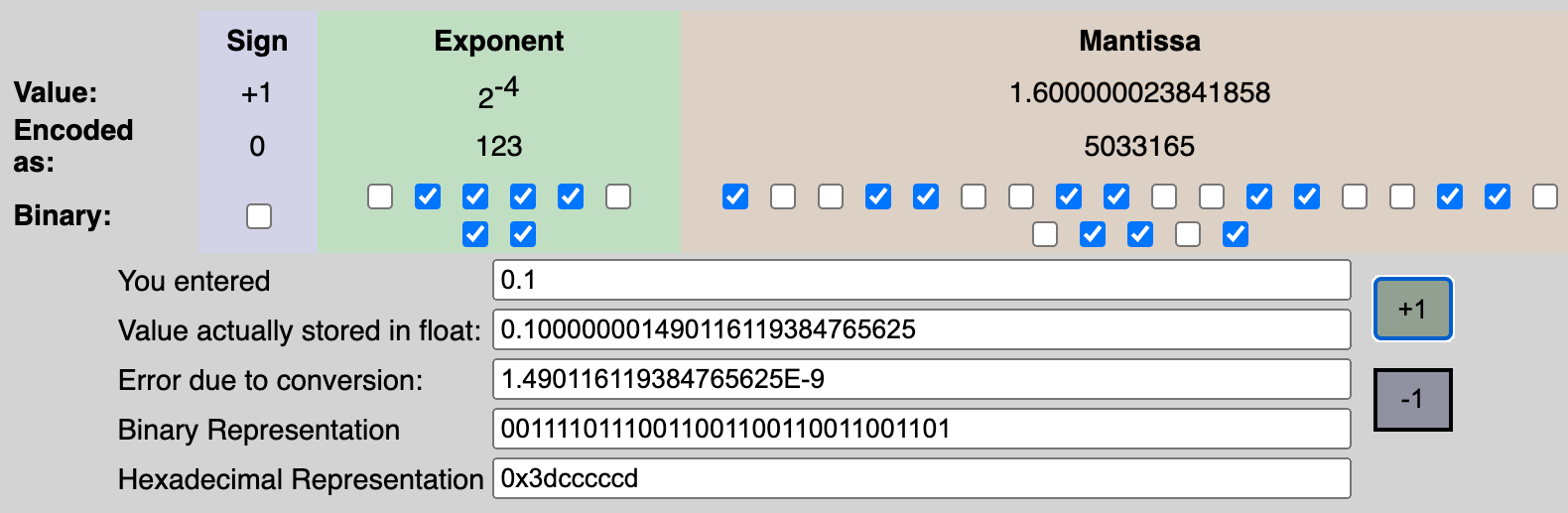}
    \caption{Internal representation of the number 0.1 in floating point, single-precision. Notice the ``error of conversion'', or round-off error.}
    \label{fig:round-off}
\end{figure}

The \emph{round-off error}, also called \emph{rounding error}, is the difference between the exact result and the result obtained using finite-precision, rounded arithmetic. Round-off errors accumulate with increasing amounts of calculation. As an example of round-off error, see the representation of the number 0.1 in figure \ref{fig:round-off}.

Notice that the number of meaningful digits that can be stored in a floating-point arithmetic, is approximately equal to: $$
\text{number of digits} \approx \frac{\text{number of bits in the mantissa}}{\log_2{(10)}}\approx \frac{\text{number of bits in the mantissa}}{3}.$$

Round-off errors accumulate with increasing amounts of calculation. If, in the course of obtaining a calculated value, one performs $N$ such arithmetic operations, one might end up having a total round-off error on the order of $\sqrt{N}\epsilon_m$ (when lucky), if the round-off errors come in randomly up or down. (The square root comes from a random-walk.) 

\subsubsection{Truncation error}
Many numerical algorithms compute ``discrete'' approximations to some desired ``continuous'' quantity. For example, an integral is evaluated numerically by computing a function at a discrete set of points, rather than at ``every'' point. In cases like this, there is an adjustable parameter, e.g., the number of points or of terms, such that the ``true'' answer is obtained only when that parameter goes to infinity. Any practical calculation is done with a finite, but sufficiently large, choice of that parameter.

The discrepancy between the true answer and the answer obtained in a practical calculation is called the \emph{truncation error}. Truncation error would persist even on a hypothetical ``perfect'' computer that had an infinitely accurate representation and no round-off error. As a general rule there is not much that a programmer can do about round-off error, other than to choose algorithms that do not magnify it unnecessarily (see discussion of “stability” below). Truncation error, on the other hand, is entirely under the programmer’s control. In fact, it is only a slight exaggeration to say that clever minimisation of truncation error is practically the entire content of the field of numerical analysis!

\subsubsection{An example: finite differentiation}
Imagine that you have a procedure which computes a function $f(x)$, and now you want to compute its derivative $f^\prime(x)$. Easy, right? The definition of the derivative, the limit as $h\to0$ of
\begin{equation}
f^{\prime}\left(x\right)\approx\frac{f\left(x+h\right)-f\left(x\right)}{h}
\label{eqn:diff}
\end{equation}
practically suggests the program: Pick a small value h; evaluate f(x + h); you probably have f(x) already evaluated, but if not, do it too; finally apply equation (\ref{eqn:diff}). What more needs to be said? Quite a lot, actually. Applied uncritically, the above procedure is almost guaranteed to produce inaccurate results. 
There are two sources of error in equation (\ref{eqn:diff}): truncation error and round-off error.
\begin{description}
\item[Truncation error.] We know that (Taylor expansion)
\[
f\left(x+h\right)=f\left(x\right)+hf^{\prime}\left(x\right)+\frac{1}{2}h^{2}f^{\prime\prime}\left(x\right)+\ldots
\]
therefore
\[
\frac{f\left(x+h\right)-f\left(x\right)}{h}=f^{\prime}+\frac{1}{2}hf^{\prime\prime}+\ldots
\]
Then, when we approximate $f^\prime$ as in equation (\ref{eqn:diff}), we make a truncation error:
\[
\varepsilon_{t}=\frac{1}{2}hf^{\prime\prime}+\ldots
\]
In this case, the truncation error is linearly proportional to $h$,
$$\varepsilon_{t}=O(h).$$
\item[The round-off error.] The round-off error has various contributions:
\begin{itemize}
\item Neither $x$ nor $x+h$ is a number with an exact binary representation; each is therefore represented with some fractional error characteristic of the machine\textquoteright s floating-point format, $\varepsilon_{m}$.
Since they are represented in the machine as rounded to the machine precision, the difference between $x+h$ and $x$ is $x\varepsilon_{m}$.
A good strategy is to choose $h$ so that $x+h$ and $x$ differ by an \emph{exact number}, for example using the following construct:
\begin{align*}
\texttt{temp} & =\texttt{x}+h\\
h & =\texttt{temp}-\texttt{x}
\end{align*}
Beware that the compiler could ``optimize-out'' these lines. Depending on the compiler and on the language, dedicated keywords can be used to prevent this.
\item The fractional accuracy with which $f$ is computed is at least
\[
\varepsilon_{r}=\varepsilon_{m}\left|\frac{f\left(x\right)}{h}\right|
\]
(but for a complicated calculation with additional sources of inaccuracy it might be larger)
\end{itemize}
\item So one has a total error
$$
\varepsilon_{\text{total}}=\varepsilon_{t}+\varepsilon_{r}=\frac{1}{2}hf^{\prime\prime}+\varepsilon_{m}\left|\frac{f\left(x\right)}{h}\right|
\label{eqn:epstotal}
$$
Equation (\ref{eqn:epstotal}) allows one to determine the optimal choice of $h$ which minimizes $\varepsilon_{\text{total}}$:
\[
h\sim\sqrt{\frac{\varepsilon_{m}f}{f^{\prime\prime}}}
\]
\end{description}
\paragraph{Remark.} Notice that, if one takes a \emph{central difference},
$$f^{\prime}\left(x\right)\approx\frac{f\left(x+h\right)-f\left(x-h\right)}{2h},$$
rather than a right difference like in equation (\ref{eqn:diff}), the calculation is more accurate, as the truncation error becomes $$\varepsilon_t = O(h^2).$$
The demonstration is left as an exercise to the reader.

\subsubsection{Underflow and overflow errors, cancellation error}
The underflow  is a condition in a computer program where the result of a calculation is a number of smaller absolute value than the computer can actually represent in memory.
Overflow is condition occurring when an operation attempts to create a numeric value that is outside of the range that can be represented with a given number of digits – either higher than the maximum or lower than the minimum representable value.

Try for instance:
$$1e100 + 1 - 1e100$$
What do you obtain? Most likely, 0. Additions and subtractions between numbers that differ in magnitude by a factor that is larger than the machine precision are likely to incur in underflow or overflow errors. This is also called \emph{cancellation} error (that sometimes can even be \emph{catastrophic} as in the example above).

Expressions like the following,
$$x^2-y^2,$$
can incur in underflow errors if $y^2\ll{x^2}$ (in particular, and more precisely, when $y^2<x^2\varepsilon_m$). Such an expression is more accurately evaluated as $$\left(x+y\right)\left(x-y\right).$$
Some cancellation could still occur, however, avoiding the squared power operation, it will be less catastrophic.

\subsubsection{Numerical stability}
Let's take the function  ``sin cardinal'' as an example. Sin cardinal, $\mathrm{sinc}(x)$, also called the ``sampling function'', is a function that arises frequently in signal processing and the theory of Fourier transforms, and it is defined as
\[
\mathrm{sinc}(x)=\begin{cases}
1 & \text{for }x=0\\
{\displaystyle \frac{\sin x}{x}} & \text{otherwise.}
\end{cases}
\]
The implementation of this function requires special attention because, when $x\to0$, numerical instabilities appear due to the division between two nearly-zero numbers. A robust implementation comes from a careful consideration of this function. 
Let's take the Taylor expansion $\mathrm{sinc}(x)$ to first order,
$$ \frac{\sin x}{x} \approx 1 - \frac{x^2}6 + \ldots $$
If we look at the right-hand side, we can appreciate the fact that in this form, when $x$ is small, the numerical instability simply disappears. The final result will differ from zero if and only if
$$ \left| -\frac{x^2}6 \right| < \varepsilon_m,$$
or, if $x$ is made explicit, if
$$ \left| x \right| < \sqrt{6\,\varepsilon_m}.$$
This leads to a robust implementation of the function \verb|sinc|:
\begin{lstlisting}
function sinc(x)
    taylor_limit = sqrt(6*epsilon);
    if abs(X) < taylor_limit
        return 1;
    endif
    return sin(x)/x;
endfunction
\end{lstlisting}

\subsection{Exact numbers}
In cases where double-, extended- or even quadruple-precision are not enough, there exist a couple of solutions to achieve higher precision and in some cases even exact results. One case use more bits of precision, from a few hundred up to thousands or even millions of bits, or symbolic calculation.

\paragraph{Symbolic calculation}
Symbolic calculation is the ``holy grail'' of exact calculations. Programs such as Maxima know the rules of math and represents data as symbols rather rounded numbers. For example: 1/3 is actually a fraction ``one divided by three''.
Even transcendental numbers like $e$ and $\pi$ are known, together with their properties, so that $e^{i\pi}$ is exactly equal to $-1$.
Symbolic math systems are invaluable tools, but they are complex, slow, and generally not appropriate for computationally intensive simulations. They should rather be used as helper tools to develop faster and more accurate algorithms. For example, to simplify expressions before they are coded in faster languages, with the aim to limit the number of operations affected by truncation and round-off errors. In the section \ref{sec:symbolic} we will expand more this topic.

\paragraph{Arbitrary precision arithmetic} When numerical calculations with precision higher than double-precision floating points, one can resolve to use dedicated libraries that can handle arbitrary, user-defined precision such as GMP, the GNU Multiple Precision Arithmetic Library for the C and C++ programming languages.

GMP is a free library for arbitrary precision arithmetic, operating on signed integers, rational numbers, and floating-point numbers. There is no practical limit to the precision except the ones implied by the available memory in the machine GMP runs on. GMP has a rich set of functions, and the functions have a regular interface.

In case one needs to perform sporadically operations in arbitrary precision, one can use the shell command \verb|bc|. See the dedicated paragraph in the following pages.

\section{Scientific programming languages}
Scientific programming languages allow one to perform numerical computations easily, using high-level concepts such as matrices, complex numbers, data processing, statistical analysis, fitting procedures. Their focus more on easiness of use and richness of the numerical toolboxes available, than on rapidity. Octave~\cite{Eaton2019} and Scilab~\cite{scilab} are excellent examples of scientific languages (and they are modeled to resemble Matlab). Also Python~\cite{van1995python} gained some popularity in this field, as explained below.

\subsection{Octave}
GNU Octave is a high-level language primarily intended for numerical computations. It is typically used for such problems as solving linear and nonlinear equations, numerical linear algebra, statistical analysis, and for performing other numerical experiments. It may also be used as a batch-oriented language for automated data processing.

The current version of Octave executes in a graphical user interface (GUI). The GUI hosts an Integrated Development Environment (IDE) which includes a code editor with syntax highlighting, built-in debugger, documentation browser, as well as the interpreter for the language itself. A command-line interface for Octave is also available.

\subsection{Python}
Python is a general-purpose programming language. Through modules such as \verb|numpy|, \verb|scipy|, and \verb|matplotlib|, the Python instruction set grows significantly to include a rich set of functions very close to those provided by Octave and Matlab.


\begin{itemize}
    \item Row Major Order: When matrix is accessed row by row (Python and C language)
    \item Column Major Order: When matrix is accessed column by column (Mathematics)
\end{itemize}

\subsection{Gnuplot}
Gnuplot is a portable command-line driven graphing utility for Linux, OS/2, MS Windows, OSX, VMS, and many other platforms. The source code is copyrighted but freely distributed (i.e., you don't have to pay for it). It was originally created to allow scientists and students to visualize mathematical functions and data interactively, but has grown to support many non-interactive uses such as web scripting. It is also used as a plotting engine by third-party applications like Octave. Gnuplot has been supported and under active development since 1986.

Not many people know that Gnuplot can also be used for fitting functions. Gnuplot uses an excellent implementation of the nonlinear least-squares (NLLS) Marquardt-Levenberg algorithm to a set of data points $(x,y)$ or $(x,y,z)$. Any user-defined variable occurring in the function body may serve as a fit parameter, but the return type of the function must be real. Here it follows an example:
\begin{lstlisting}
     f(x) = a*x**2 + b*x + c
     g(x,y) = a*x**2 + b*y**2 + c*x*y
     FIT_LIMIT = 1e-6
     fit f(x) 'measured.dat' via 'start.par'
     fit f(x) 'measured.dat' using 3:($7-5) via 'start.par'
     fit f(x) './data/trash.dat' using 1:2:3 via a, b, c
     fit g(x,y) 'surface.dat' using 1:2:3:(1) via a, b, c
\end{lstlisting}

\section{Symbolic computation\label{sec:symbolic}}

Computer algebra, also called symbolic computation or algebraic computation, is a scientific area that refers to algorithms and procedures for manipulating mathematical expressions and other mathematical objects. Computer algebra is generally considered as a distinct field of scientific computing, because scientific computing is usually based on numerical computation with approximate floating point numbers (truncation error!), while symbolic computation emphasizes exact computation with expressions containing variables that have no given value and are manipulated as symbols.

Computer algebra and symbolic calculation can also be used to optimize expressions prior to their implementation in other languages like Octave, Python, C, and C++, in order to minimize truncation and round-off errors.

\subsection{Maxima}
\begin{figure}
    \centering
    \includegraphics[width=0.9\textwidth]{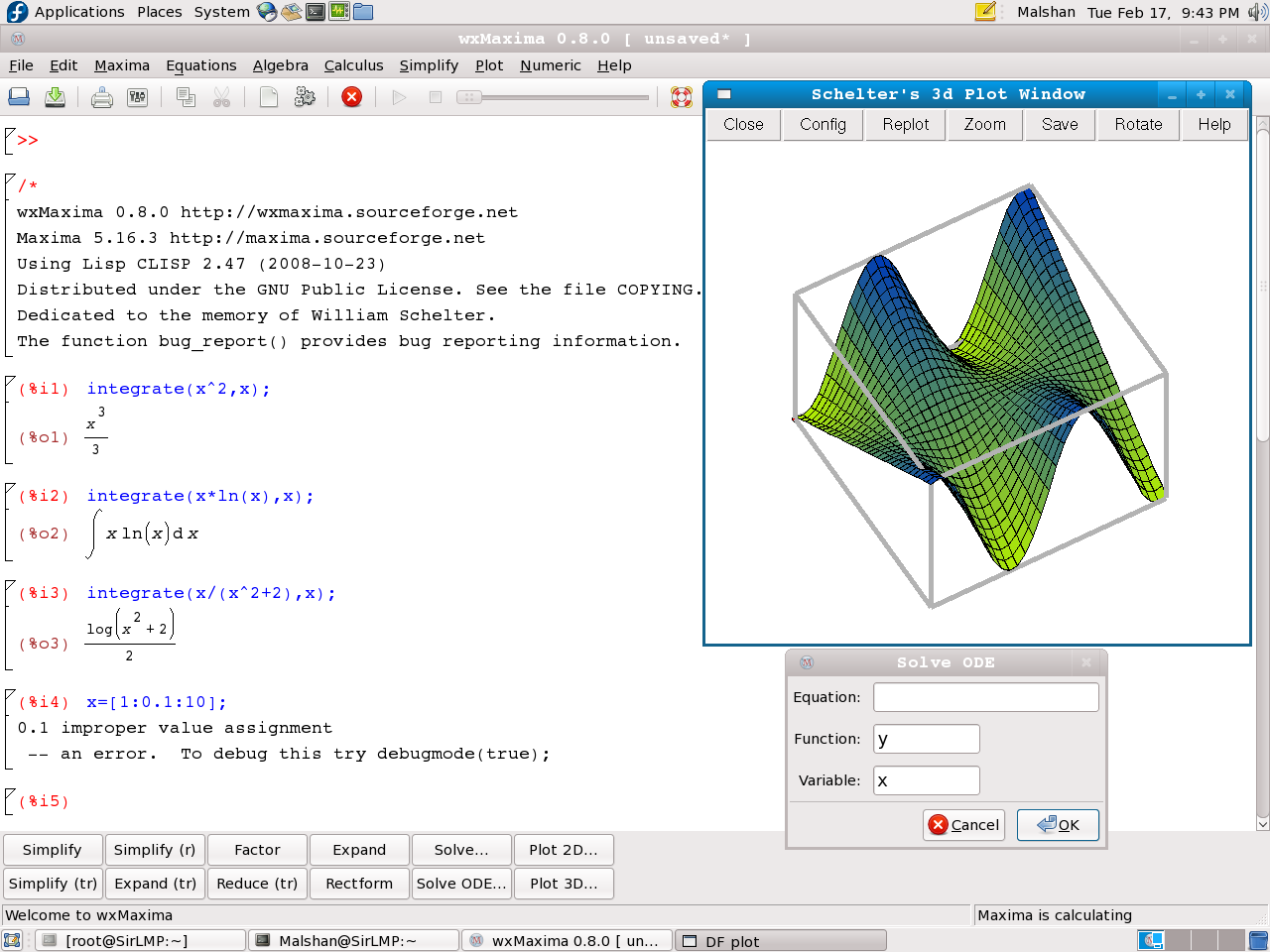}
    \caption{A screenshot of wxMaxima running on Linux, with Gnome interface.}
    \label{fig:wxMaxima}
\end{figure}

Maxima is a computer algebra system with a long history. It is based on a 1982 version of Macsyma, it is written in Common Lisp and runs on all POSIX platforms such as macOS, Unix, BSD, and Linux, as well as under Microsoft Windows and Android. It is free software released under the terms of the GNU General Public License (GPL). An excellent front end for Maxima is wxMaxima~\cite{wxMaxima}, see figure \ref{fig:wxMaxima}.

Maxima is a system for the manipulation of symbolic and numerical expressions, including differentiation, integration, Taylor series, Laplace transforms, ordinary differential equations, systems of linear equations, polynomials, sets, lists, vectors, matrices and tensors.
Maxima yields high precision numerical results by using exact fractions, arbitrary-precision integers and variable-precision floating-point numbers. Maxima can plot functions and data in two and three dimensions.

Besides the fact of being free software and open-source, a great advantage of Maxima versus its more popular competitors (e.g., Mathematica), is that its syntax and outputs are compatible with languages like Octave, Python, C, and C++. This makes it possible to develop a complex calculation (like the harmonic oscillator, below), then copy it \& paste it directly into your simulation code. 

\subsubsection{Maxima: 1-D harmonic oscillator}
As an example of Maxima usage, we resolve the 2$^\text{nd}$-order differential equation of a 1-D harmonic oscillator, where we express as $\beta$ the restraining focusing force. Of course, there is a connection between $\beta$, and the Twiss parameter $\beta$, known as $\beta$ function. We start with declaring $\beta$ as a positive constant.

\noindent
\begin{minipage}[t]{4em}\color{red}
(\% i1) 
\end{minipage}
\begin{minipage}[t]{\textwidth}\color{blue}
declare(beta, constant)\$
\end{minipage}

\noindent
\begin{minipage}[t]{4em}\color{red}
(\% i2)
\end{minipage}
\begin{minipage}[t]{\textwidth}\color{blue}
assume(beta\ensuremath{>}0);
\end{minipage}
(\% o2) \centerline{$\displaystyle[0<\beta]\mbox{}$}
~\\Follows the equation of motion. Notice the symbol `` ' '' : it tells Maxima to defer the evaluation of the derivative to a later moment.
\\\noindent
\begin{minipage}[t]{4em}\color{red}
(\% i3)
\end{minipage}
\begin{minipage}[t]{\textwidth}\color{blue}
eqn\_1:  'diff(x(s), s, 2) + x(s)/beta**2 = 0;
\end{minipage}
eqn\_1: \centerline{$\displaystyle
\frac{{{d}^{2}}}{d {{s}^{2}}} {x}(s)+\frac{{x}(s)}{{{\beta}^{2}}}=0\mbox{}$}
~\\Let's set the initial conditions
\\\noindent
\begin{minipage}[t]{4em}\color{red}
(\% i4)
\end{minipage}
\begin{minipage}[t]{\textwidth}\color{blue}
atvalue(x(s), s=0, x0)\$
\end{minipage}
\\\noindent
\begin{minipage}[t]{4em}\color{red}
(\% i5)
\end{minipage}
\begin{minipage}[t]{\textwidth}\color{blue}
atvalue('diff(x(s),s,1), s=0, xp0)\$
\end{minipage}
~\\Solution

\noindent
\begin{minipage}[t]{4em}\color{red}
(\% i6)
\end{minipage}
\begin{minipage}[t]{\textwidth}\color{blue}
desolve(eqn\_1, x(s));
\end{minipage}
(\% o6) \centerline{$
\displaystyle
{x}(s)=\frac{{{\beta}^{3}} \sin{\left( \frac{s}{\beta}\right) } \mathit{xp0}+{{\beta}^{2}} \cos{\left( \frac{s}{\beta}\right) } \mathit{x0}}{{{\beta}^{2}}}\mbox{}
$}
\noindent
\begin{minipage}[t]{4em}\color{red}
(\% i7)
\end{minipage}
\begin{minipage}[t]{\textwidth}\color{blue}
ratsimp(\%);
\end{minipage}
(\% o7) \centerline{$
\displaystyle
{x}(s)=\beta \sin{\left( \frac{s}{\beta}\right) } \mathit{xp0}+\cos{\left( \frac{s}{\beta}\right) } \mathit{x0}\mbox{}
$}
\noindent
\begin{minipage}[t]{4em}\color{red}
(\% i8)
\end{minipage}
\begin{minipage}[t]{\textwidth}\color{blue}
diff(\%,s);
\end{minipage}
(\% o8) \centerline{$
\frac{d}{d s} {x}(s)=\cos{\left( \frac{s}{\beta}\right) } \mathit{xp0}-\frac{\sin{\left( \frac{s}{\beta}\right) } \mathit{x0}}{\beta}\mbox{}
$}


\subsection{Symbolic computation in Python and Octave}
Symbolic computations can also be performed within Octave and Python. This adds the possibility to perform basic symbolic computations, including common Computer Algebra System tools such as algebraic operations, calculus, equation solving, Fourier and Laplace transforms, variable precision arithmetic and other features, in scripts.

\subsubsection{Octave ``symbolic'' package}
Here follows an example of Octave symbolic:
\begin{lstlisting}
% Load the symbolic package
pkg load symbolic

% This is just a formula to start with, have fun and change it if you want to.
f = @(x) x.^2 + 3*x - 1 + 5*x.*sin(x);

% These next lines take the Anonymous function into a symbolic formula
syms x;
ff = f(x);

% Now we can calculate the derivative of the function
ffd = diff(ff, x);

% and convert it back to an Anonymous function
df = function_handle(ffd)
\end{lstlisting}

\subsubsection{Sympy}
The Python package Sympy offers similar capabilities.
\begin{lstlisting}
>>> from sympy import *
>>> x = symbols('x')
>>> simplify(sin(x)**2 + cos(x)**2)
1
\end{lstlisting}

\section{High-performance computing: C and C++}
For intensive computations, no scripting language can beat the speed of compiled languages such as C and C++. The C++ version of a simulation can be hundreds or even thousands of times faster than the equivalent script written in Python or Octave.

The price to pay for such high speed is the limited expressiveness of the C and C++ languages and their lower level. Programming in C and C++ is closer to programming in the assembly language understood by the CPU than any of the aforementioned languages. Writing code in C is nearly equivalent to writing in assembly directly. 

In other words, C and C++ are two low-level languages. Which is the reason why, on the one hand, they are regarded as ``difficult'' languages, but on the other hand it is their greatest power, and the reason for their unbeatable speed. Through pointers, for example, one can directly access memory areas and process data at light speed. Which unfortunately is the same speed at which one can heavily mess up things, for example by mistakenly accessing memory belonging to other processes.

Being the C language well established and standardized, despite low-level, a number of libraries exist to provide advanced numerical tools. One such library is the GNU Scientific Library (GSL), which implements a large variety of high-quality routines spanning nearly all aspects of numerical computing. The library provides a wide range of mathematical routines such as random number generators, special functions and least-squares fitting. There are over 1000 functions in total. GSL is free software under the GNU General Public License.

The object-oriented features of C++ come to help with the possibility to write and utilize complex ``objects'' (or classes) that can accomplish complex tasks in a safe manner, without sacrificing performance. This effectively means that, in C++, one can ``customize'' the language adding high-level objects and types to serve virtually any purpose. Of course, mathematical objects and physics laws find a perfect realisation in C++ classes, which makes C++ an ideal language for high-performance computing.

\subsection{A word about C matrices}
The C and C++ language specifications (as well as Python) state that arrays are laid out in memory in a row-major order: the elements of the first row are laid out consecutively in memory, followed by the elements of the second row, and so on. The way we access a matrix impacts the performance of our code. See this example:
\begin{lstlisting}
// C program showing time difference
// in row major and column major access
#include <stdio.h>
#include <time.h>
 
// taking MAX 10000 so that time difference
// can be shown
#define MAX 10000
 
int arr[MAX][MAX] = {0};
 
void rowMajor() {
  int i, j;
  // accessing element row wise
  for (i = 0; i < MAX; i++) {
    for (j = 0; j < MAX; j++) {
      arr[i][j]++;
    }
  }
}
 
void colMajor() {
  int i, j;
  // accessing element column wise
  for (i = 0; i < MAX; i++) {
    for (j = 0; j < MAX; j++) {
      arr[j][i]++;
    }
  }
}
 
// driver code
int main() {
  int i, j;
  // Time taken by row major order
  clock_t t = clock();
  rowMajor();
  t = clock() - t;
  printf("Row major access time: %f s\n", t / (float)CLOCKS_PER_SEC);
 
  // Time taken by column major order
  t = clock();
  colMajor();
  t = clock() - t;
  printf("Column major access time: %f s\n", t / (float)CLOCKS_PER_SEC);
  return 0;
}
\end{lstlisting}
The output is:
\begin{lstlisting}
Row major access time: 0.492000 s
Column major access time: 1.621000 s
\end{lstlisting}
Notice that Octave use Matlab column-major order, which might lead to some confusion sometimes.

\subsection{Linear algebra}
\subsubsection{BLAS and LAPACK}
LAPACK~\cite{LAPACK} is a library of Fortran 77 subroutines for solving the most commonly occurring problems in numerical linear algebra. It has been designed to be efficient on a wide range of modern high-performance computers. The name LAPACK is an acronym for Linear Algebra PACKage.

LAPACK routines are written so that as much as possible of the computation is performed by calls to the Basic Linear Algebra Subprograms (BLAS)~\cite{BLAS}. BLAS defines a set of fundamental operations on vectors and matrices which can be used to create optimized higher-level linear algebra functionality. Highly efficient machine-specific implementations of the BLAS are available for many modern high-performance computers. The BLAS enable LAPACK routines to achieve high performance with portable code. The BLAS are not strictly speaking part of LAPACK, but Fortran 77 code for the BLAS is distributed with LAPACK, or can be obtained separately from netlib. 

The library provides a low-level layer which corresponds directly to the C-language BLAS standard, referred to here as “CBLAS”, and a higher-level interface for operations on GSL vectors and matrices. Users who are interested in simple operations on GSL vector and matrix objects should use the high-level layer described in this chapter.

\subsubsection{Sparse matrices}
Matrices which are populated primarily with zeros and contain only a few non-zero elements are called ``sparse matrices''. Sparse matrices often appear in the solution of partial differential equations, like Poisson solvers. It is beneficial to use specialized data structures and algorithms for storing and working with sparse matrices, since dense matrix algorithms and structures can be prohibitively slow and use huge amounts of memory when applied to sparse matrices. Good implementations of sparse matrices come with GSL (see below), and SuiteSparse \href{https://people.engr.tamu.edu/davis/welcome.html}{[link]}.

\subsection{C++ template libraries}

Templates are one of the most interesting and precious constructs of C++. operate with generic types. Templates allow the creation of functions and classes whose functionality can be adapted to more than one type or class without repeating the entire code for each type. Due to the implementation of templates in modern compilers, which enables significant optimisation at compilation time, they produce extremely efficient and fast code.
For instance, one can imagine a C++ class dedicated to $2\times 2$ matrices,
\begin{lstlisting}
// Example of simple template class
template <class T>
class Matrix22 {
    T r11, r12, r21, r22;
public:
    friend Matrix22 operator(const Matrix22 &a, const Matrix22 &b )
    {
        Matrix22 r;
        r.r11 = b.r21*a.r12+b.r11*a.r11;
        r.r12 = b.r22*a.r12+b.r12*a.r11;
        r.r21 = b.r21*a.r22+b.r11*a.r21;
        r.r22 = b.r22*a.r22+b.r12*a.r21;
        return r;
    }
    T determinant() const { return r11*r22 - r21*r12; }
};
\end{lstlisting}
To be used to allocate and operate on matrices of different types:
\begin{lstlisting}
// Declare three 2x2 matrices of different type
Matrix22<int> int_matrix;
Matrix22<double> dbl_matrix;
Matrix22<std::complex<double>> cmp_matrix;
    
// Print out the determinants
std::cout << int_matrix.determinant() << std::endl;
std::cout << dbl_matrix.determinant() << std::endl;
std::cout << cmp_matrix.determinant() << std::endl;
\end{lstlisting}
In short, templates are used to teach the compiler the archetype of a function, regardless of its actual type. In our example, our template class \verb|Matrix22| described what a $2\times2$ matrix is, regardless of the actual ``content'' that will constitute the matrix itself. Oftentimes, templates let our code do more than we even imagine or planned them for. For instance, one could imagine to define a \verb|Matrix22<Matrix22>| which well represents a $4\times4$ matrix. The definition of determinant would still apply.

Template classes are extremely useful to define generic containers, such as lists, queues, vectors of elements. Their abstraction is their strength, together with the high level of optimisation that the C++ compiler can apply to the code. Template classes can also be used to define abstract functions or algorithms, as well as service classes like smart pointers.

\paragraph{C++ Standard Library}

The C++ language comes with its ``Standard Library'' which consists of a large collection of generic containers, functions to use and to manipulate such containers, generic strings and streams (including interactive and file I/O), etc. It is based upon conventions introduced by the Standard Template Library (STL), and has been influenced by research in generic programming and developers of the STL. All features of the C++ Standard Library are declared within the \verb|std| namespace.
Useful container classes are for example \verb|std::array<T>|, \verb|std::vector<T>|, \verb|std::list<T>|, \verb|std::deque<T>|. To create dictionaries and associations between types, there exist associative containers like \verb|std::set<K>|, \verb|std::map<K,T>|. 

The C++ standard library also offers efficient implementation of several algorithms, e.g., sorting algorithms, see for instance \verb|std::qsort|, which provides a very fast implementation of the quicksort algorithm.

\paragraph{BOOST - \href{https://www.boost.org/}{[link]}}
Boost is a set of C++ libraries that provides support for tasks and structures such as linear algebra, pseudorandom number generation, multithreading, image processing, regular expressions, and unit testing. It contains 164 individual libraries (as of version 1.76). Boost is meant to build on top of STL and many of the libraries are slated to become part of the standard library eventually. Boost libraries are generally less mature and less standard than STL.

\paragraph{Armadillo - \href{http://arma.sourceforge.net/}{[link]}}
Armadillo is a high quality linear algebra library for the C++ language, aiming towards a good balance between speed and ease of use. It provides high-level syntax and functionality deliberately similar to Matlab and Octave. Useful for algorithm development directly in C++, or quick conversion of research code into production environments. Armadillo provides efficient classes for vectors, matrices and cubes; dense and sparse matrices are supported; integer, floating point and complex numbers are supported. A sophisticated expression evaluator (based on template meta-programming) automatically combines several operations to increase speed and efficiency. Various matrix decompositions (eigen, SVD, QR, etc) are provided through integration with LAPACK, or one of its high performance drop-in replacements (eg. MKL or OpenBLAS). Armadillo can automatically use OpenMP multi-threading (parallelisation) to speed up computationally expensive operations. Distributed under the permissive Apache 2.0 license, useful for both open-source and proprietary (closed-source) software. Armadillo can be used for machine learning, pattern recognition, computer vision, signal processing, bioinformatics, statistics, finance, etc.

\paragraph{Eigen - \href{https://eigen.tuxfamily.org/}{[link]}}
Eigen is a(nother) C++ template library for linear algebra: matrices, vectors, numerical solvers, and related algorithms. Eigen is implemented using the expression templates metaprogramming technique, meaning it builds expression trees at compile time and generates custom code to evaluate these. Using expression templates and a cost model of floating point operations, the library performs its own loop unrolling and vectorisation. This guarantees excellent execution speed at run time. Eigen itself can provide BLAS and a subset of LAPACK interfaces.

\subsection{Random number generation}

Random number generators are widely using in numerical physics. They are at the base of each Monte Carlo technique, and are often the only practical way to evaluate difficult integrals or to sample random variables governed by complicated probability density functions.

\subsubsection{Pseudorandom numbers}

It might seem impossible to produce random numbers through deterministic algorithms. Nevertheless, computer ``random number generators'' are in common use.  Oftentimes, computer-generated sequences are in fact called pseudo-random, while the word random is reserved for the output of an intrinsically random physical process, like the elapsed time between clicks of a Geiger counter placed next to a sample of some radioactive element. Entire books have been dedicated to this topic, most notably~\cite{Knuth1998}.

A working, though imprecise, definition of randomness in the context of computer-generated sequences, is to say that the deterministic program that produces a random sequence should be different from, and --in all measurable respects-- statistically uncorrelated with, the computer program that uses its output.
In other words, any two different random number generators ought to produce statistically the same results when coupled to your particular applications program. If they don’t, then at least one of them is not (from your point of view) a good generator.

All of the algorithms produce a periodic sequence of numbers, and to obtain effectively random values, one must not use more than a small subset of a single period. The quality of a random number generator is defined on the randomness of each sequence, and on the length of the period. There exist tens, if not hundreds, of random number generators.

The performance of the generators can be investigated with tests such as DIEHARD \cite{MARSAGLIA1985}  or TestU01~\cite{Zyla2020a}. Good random number generators ought to pass these tests; or at least the user had better be aware of any that they fail, so that he or she will be able to judge whether they are relevant to the case at hand. Good quality uniformly distributed random numbers, lying within a specified range (typically 0 to 1), are an essential building block for any sort of stochastic modeling or Monte Carlo computer work as they can be used to generate any other distribution. 

A short description of the most common generators can be found in the Gnu Scientific Library documentation~\cite{Galassi2009}, also on-line.

\subsubsection{Quasirandom numbers}

Quasi-random sequences are sequences that progressively cover a $N$-dimensional space with a set of points that are uniformly distributed. Quasi-random sequences are also known as low-discrepancy sequences. The quasi-random sequence generators use an interface that is similar to the interface for random number generators, except that seeding is not required—each generator produces a single sequence.

Unlike the pseudo-random sequences, quasi-random sequences fail many statistical tests for randomness. Approximating true randomness, however, is not their goal. Quasi-random sequences seek to fill space uniformly, and to do so in such a way that initial segments approximate this behavior up to a specified density.

\begin{figure}
    \centering
    \includegraphics[width=0.6\textwidth]{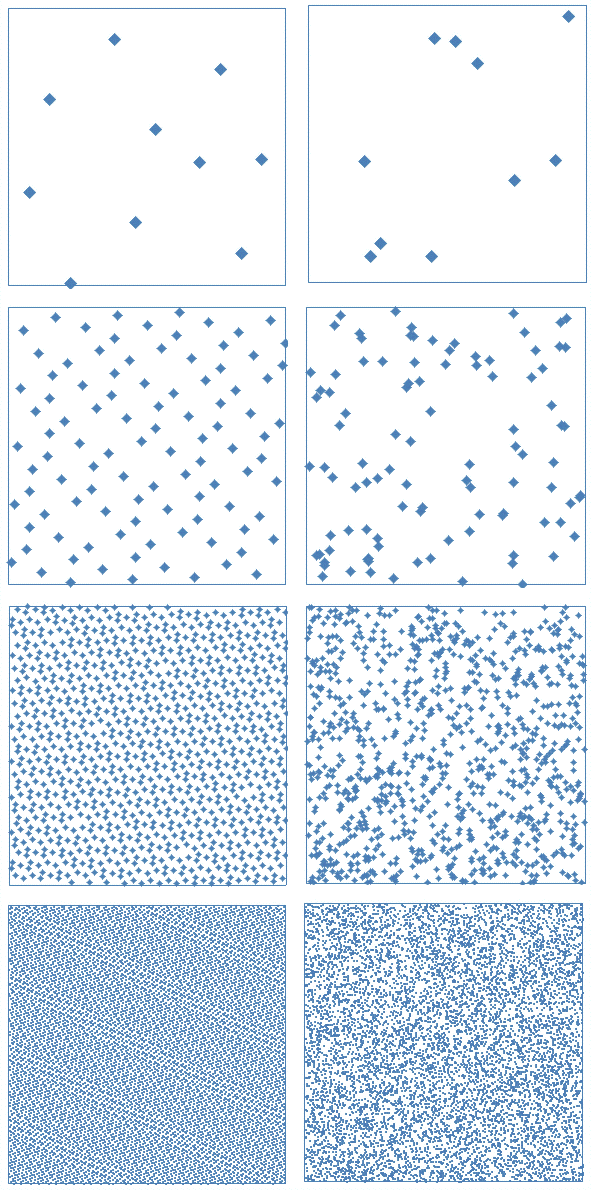}
    \caption{Coverage of the unit square. Left for quasi-random numbers. Right for pseudo-random numbers. From top to bottom. 10, 100, 1000, 10000 points. (Source: Wikipedia)}
    \label{fig:quasirandom}
\end{figure}

\subsection{Numerical libraries}
\subsubsection{GSL - The GNU Scientific Library}

The GNU Scientific Library (GSL) is an excellent numerical library for C and C++ programmers. It provides a wide range of mathematical routines such as random number generators, special functions and least-squares fitting. There are over 1000 functions in total with an extensive test suite. THe GSL is written in C, and can be called from C, C++, and even Fortran programs.
The GSL it's free software under the GNU General Public License.

\subsubsection{The origins: the NAG library}
The NAG Numerical Library is the mainstream, commercial, cumbersome precursor of GSL. It's a library of numerical analysis routines, containing nearly 2,000 mathematical and statistical algorithms. Areas covered by the library include linear algebra, optimisation, quadrature, the solution of ordinary and partial differential equations, regression analysis, and time series analysis. This library is developed and sold by The Numerical Algorithms Group. 

The original version of the NAG Library was written in Algol 60 and Fortran. It contained 98 user-callable routines, and was released for the ICL 1906A and 1906S machines on October 1, 1971. The first partially vectorized implementation of the NAG Fortran Library for the Cray-1 was released in 1983, while the first release of the NAG Parallel Library (which is specially designed for distributed memory parallel computer architectures) was in the early 1990s. Mark 1 of the NAG C Library was released in 1990. In 1992, the Library incorporated LAPACK routines for the first time; NAG had been a collaborator in the LAPACK project since 1987. 

\subsection{Parallelism}

Over the last decade, CPUs have progressively switched towards more and more parallel computing. Parallelism, that is splitting a problem in tasks that can be performed simultaneously, can dramatically reduce the  computation speed by order of magnitudes. Several models of parallelisation exist, depending on the problem being studied. There is not a general parallelisation strategy, and often the best solution for a given problem does not work with another. It also depends on the hardware we intend to run our software on. 
For instance, parallel programming technologies such as MPI are used in a distributed computing environment (dedicated clusters of computer connected by fast links to be used), while multi-threads programs are limited to a single computer system: all threads within a process share the same address space. In some specific cases, like for instance lattice QCD calculations, the physical implementation of the cluster matches the lattice structure of the problem in study: the cluster's nodes are physically connected to their first neighbors to reproduce the lattice structure being studied.

In this section we will give a few examples that should illustrate the variety of solutions in accelerator-physics problems.

\subsubsection{Embarrassingly parallel problems}
This is the class of problems that gives more satisfaction and that better benefits from parallelism. Embarrassingly parallel problems are those where a large number of tasks need to be performed, with each single task being completely independent of the others. In this case, all tasks can in principle be performed simultaneously, and the speed up factor is proportional to the number of processes that one can run in parallel. 

The simulation of accelerator imperfections for instance, where hundreds of different random misalignment configurations are tested in order the sample the response of a system to random installation errors, provides a perfect example of embarrassingly parallel problem. 

In this case, no modifications are needed to the simulation code -which can even be sequential and run on a single core-, and all the random configurations can be spawn to hundreds of different computer and run independently and simultaneously. Practically, this can be achieved using software solutions called ``job schedulers'', like for instance HTCondor (which is in use at CERN), that help running computations on a pool of multiple computers (or a ``farm'' of computers).

\subsubsection{MPI}
For computational problems that can be solved with parallel algorithms, several approaches exist. One such approaches, typical of massively parallel problems, consists in designing and writing to run on clusters of computers. In this case, the software must be conceived, designed, and written to function on parallel computing architectures (e.g., a cluster of computers, called nodes, connected between each other using high-speed low-latency connections). 

A standardized and portable solution to handle message-passing (and data sharing) between nodes of a cluster is the ``Message-Passing Interface'' (MPI). MPI is not a library in itself. The MPI is a standard that defines the syntax and semantics of a set of routines useful to a wide range of users writing portable message-passing programs in C, C++, and Fortran. There exist several open-source implementations of MPI, which fostered the development of a parallel software industry, and encouraged development of portable and scalable large-scale parallel applications. Two well-established MPI implementations are ``Open MPI'' and ``MPICH''.

In spite of the unequivocal advantage of controlling from a single code a cluster of computers, writing MPI code requires a considerable effort. A code must be \emph{conceived} and \emph{designed} to run on MPI from the beginning. Adapting an existing code to run on MPI is a nearly impossible task. To give an example, follows an example of ``Hello world'' written in C using MPI:
\begin{lstlisting}
#include <mpi.h>
#include <stdio.h>

int main()
{
    // Initialize the MPI environment
    MPI_Init(NULL, NULL);

    // Get the number of processes
    int world_size;
    MPI_Comm_size(MPI_COMM_WORLD, &world_size);

    // Get the rank of the process
    int world_rank;
    MPI_Comm_rank(MPI_COMM_WORLD, &world_rank);

    // Get the name of the processor
    char processor_name[MPI_MAX_PROCESSOR_NAME];
    int name_len;
    MPI_Get_processor_name(processor_name, &name_len);

    // Print off a hello world message
    printf("Hello world from processor %s, rank %d out of %d processors\n",
           processor_name, world_rank, world_size);

    // Finalize the MPI environment.
    MPI_Finalize();
}
\end{lstlisting}

\subsubsection{OpenMP}
Hacking an existing code to make it parallel, it's best done with OpenMP. OpenMP is a programming interface that supports multi-platform shared-memory multiprocessing programming in C, C++, and Fortran. In simpler words, it makes programs run in parallel on multi-cores computers, exploiting the multi-threaded architecture of modern CPUs.
OpenMP is best understood by looking at an example:
\begin{lstlisting}
int main()
{
    int a[100000];

    #pragma omp parallel for
    for (int i = 0; i < 100000; i++) {
        a[i] = 2 * i;
    }

    return 0;
}
\end{lstlisting}
In this code, the pragma \verb|omp parallel| is used to fork additional threads and carry out the work enclosed in the construct in parallel. This specific pragma applies to loops that are ``embarrassingly parallel'' (each iteration is independent of the others), however more sophisticated schemes exist to handle more complex cases.

\subsubsection{C++ threads}
When designing and writing a code from scratch, with a multi-core / multi-threaded architecture in mind (like basically \emph{any} computer today, and even a smart phone), the best solution to handle parallelism is to use functions and constructs offered by the language of choice. This ensures portability across systems and better integration within the language. Modern C++, since the standard version C++11, offers a set of classes to handle parallelism, synchronisation, and data exchange between threads. This revolves around the class \verb|thread|, defined in the include file \verb|<thread>|.

For instance, in C++, one can easily implement a generic ``parallel for'' through the use of lambda functions and ``functors''. Here is the header file that implements it:
\begin{lstlisting}
#ifndef parallel_for_hh
#define parallel_for_hh

#include <algorithm>
#include <cstddef>
#include <vector>
#include <thread>

template <typename Function>
size_t parallel_for(size_t Nthreads, int begin, int end, Function func )
{
  const int size = end - begin;
  // func must be in the form: func(thread, start, end)
  if (size<Nthreads)
    Nthreads = size;
  if (Nthreads) {
    std::vector<std::thread> threads(Nthreads-1);
    for (size_t thread=1; thread<Nthreads; thread++) {
      int from = begin + thread * size / Nthreads;
  	  int to = begin + (thread+1) * size / Nthreads;
  	  threads[thread-1] = std::thread(func, thread, from, to);
    }
    func(0, begin, begin + size / Nthreads);
    for (auto &t: threads)
      t.join();
  }
  return Nthreads; // returns the effective number of threads used  
}

#endif /* parallel_for_hh */
\end{lstlisting}
The syntax is considerably more verbose than the equivalent offered by OpenMP. Nevertheless, the advantage of \verb|std::thread| is that it's part of the standard C++ language. Therefore, its integration within the language is much deeper and allows one to design custom, sophisticated, and optimized parallelisation schemes, with way more control that OpenMP can possibly offer. Mimicking the OpenMP example, it can be used as follows:
\begin{lstlisting}
#include "parallel_for.hh"

int main()
{
  std::vector<int> a(100000);

  auto do_something = [&] (size_t thread, int start, int end ) {
    for (int i = start; i < end; i++) {
      a[i] = 2 * i;
    }
  };

  size_t max_number_of_threads = std::thread::hardware_concurrency();
  parallel_for(max_number_of_threads, 0, a.size(), do_something);
  return 0;
}
\end{lstlisting}

\subsubsection{C threads}
The C language does not support multi-threading natively. To achieve parallelism in C one can use the Portable Operating System Interface (POSIX) set of libraries. POSIX is a family of standards specified by the IEEE Computer Society for maintaining compatibility between operating systems.

POSIX defines the both system- and user-level application programming interfaces. The POSIX thread libraries are a standards based thread API for C/C++. It allows one to spawn a new concurrent process flow. It is most effective on multi-processor or multi-core systems where the process flow can be scheduled to run on another processor thus gaining speed through parallel or distributed processing. 

The use of POSIX is quite complicated, and a detailed explanation goes beyond the scope of these lectures. Easier solutions exist using Octave, Python, C++, or even just including a rationalisation of the simulation procedure and some useful shell commands (See for example to use of FIFOs, below.)

\subsubsection{GPU Computing: OpenCL / CUDA}
The advent of powerful graphics cards has opened a new chapter in the history of high-performance computing. GPU computing is the use of a GPU (graphics processing unit) as a co-processor to accelerate general-purpose scientific and engineering computing.

The GPU accelerates applications running on the CPU by offloading some of the compute-intensive and time consuming portions of the code. The rest of the application still runs on the CPU. From a user's perspective, the application runs faster because it's using the massively parallel processing power of the GPU to boost performance. This is known as "heterogeneous" or "hybrid" computing.

The most striking difference between CPUs and GPUs is the number of available cores. High-end CPUs like Intel Xeon processors feature a number of cores up to 30s. In nVidia’s and AMD's current generation of high end GPUs has nearly 4,000 cores.
Two solutions exist to access the enormous computational power of GPUs, OpenCL and CUDA.
\begin{description}
\item[OpenCL] OpenCL (Open Computing Language) is a framework for writing programs that execute across heterogeneous platforms consisting of central processing units (CPUs), graphics processing units (GPUs), digital signal processors (DSPs), field-programmable gate arrays (FPGAs) and other processors or hardware accelerators. OpenCL specifies directives for programming these devices and APIs (application programming interfaces) to control the platform and execute programs on the compute devices. OpenCL provides a standard interface for parallel computing using task- and data-based parallelism.

OpenCL is an open standard maintained by the non-profit technology consortium Khronos Group. Conformant implementations are available from Altera, AMD, Apple, ARM, Creative, IBM, Imagination, Intel, Nvidia, Qualcomm, Samsung, Vivante, Xilinx, and ZiiLABS.
\item[CUDA] (an acronym for Compute Unified Device Architecture) is a proprietary model created by Nvidia to program Nvidia GPUs for general purpose processing. The CUDA platform is a software layer that gives direct access to the GPU's virtual instruction set and parallel computational elements, for the execution of compute kernels.
\end{description}

The challenge is that GPU computing requires the use of graphics programming languages like OpenGL and CUDA to program the GPU. And one has to rewrite from scratch scientific applications.

\paragraph{Parallel Octave and Python}
Octave and Python can provide limited parallelism through a dedicated package and a module. In Octave, an easy solution is to use the \verb|parallel| package, which is available from the Octave Sourceforge website. We refer to the online documentation of this package for more details.

In Python, the multiprocessing module is used to run independent parallel processes by using subprocesses (instead of threads). It allows one to leverage multiple processors on a machine, which means, the processes can be run in completely separate memory locations.

\subsubsection{Parallelism in the shell, using FIFOs}
A certain degree of parallelism can also be obtained without even using parallel codes, just using the Linux command line wisely.

Imagine we are simulating the two arms of electron-positron linear collider, and intend to simulate the collision once the two bunches arrive at the interaction point (IP). It is clear that the tracking of the two bunches along their respective linacs can be performed simultaneously (just like in the real world!). Only once the two bunches reach the IP they can be passed to the beam-beam simulation code.

A convenient and elegant solution to perform these operations and synchronize them, without modifying a single line of code, is to make use of the powerful ``named pipes'', of FIFO. The term ``FIFO'' refers to its first-in, first-out functioning mode, and it's a mechanism, in Unix and Linux, used to enable inter-process communication within the same machine. FIFOs allow two-way communication, using a special ``file'' as a ``meeting point'' between two processes.

So, here's an example of creating a named pipe.
\begin{lstlisting}
$ mkfifo mypipe
$ ls -l mypipe
prw-r-----. 1 myself staff 0 Jan 31 13:59 mypipe
\end{lstlisting}
Notice the special file type designation of "p" and the file length of zero. You can write to a named pipe by redirecting output to it and the length will still be zero.
\begin{lstlisting}
$ echo "Can you read this?" > mypipe
$ ls -l mypipe
prw-r-----. 1 myself staff 0 Jan 31 13:59 mypipe
\end{lstlisting}
So far, so good, but hit return and nothing much happens.
\begin{lstlisting}
$ echo "Can you read this?" > mypipe
\end{lstlisting}
While it might not be obvious, your text has entered into the pipe, but you're still peeking into the input end of it. You or someone else may be sitting at the output end and be ready to read the data that's being poured into the pipe, now waiting for it to be read.
\begin{lstlisting}
$ cat mypipe
Can you read this?
\end{lstlisting}
Once read, the contents of the pipe are gone.

Now, the tracking along the linacs has, as ultimate result, the task to leave two files on disk, say \verb|electrons.dat| and \verb|positrons.dat|, containing the two bunches at the IP. The beam-beam code will read these files as inputs and compute the luminosity. 

If we create two FIFOs named \verb|electrons.dat| and \verb|positrons.dat| our simulation codes with automatically use these FIFOs to send the electron and the positron bunches from the linacs to the beam-beam at the IP, without writing a single byte on disk.

Let's see how to do it. First, let's create the FIFOs:
\begin{lstlisting}
$ mkfifo electrons.dat
$ mkfifo positrons.dat
\end{lstlisting}
Then, we run our simulation (notice the `\verb|&|`, to run the two tracking simultaneously):
\begin{lstlisting}
$ track electron_linac.m > electrons.dat & 
$ track positron_linac.m > positrons.dat &
$ beam_beam electrons.dat positrons.dat > luminosity.dat
\end{lstlisting}
The beam-beam simulation will diligently wait for the bunches to traverse the linacs before it starts computing the luminosity.

\subsection{Advanced programming}

\subsubsection{Intrinsics}
For those who try to squeeze CPU cycle
The key for speed, in modern CPUs, is vectorisation. By explicit vectorisation one can access specific instruction sets like MMX, SSE, SSE2, SSE3, AVX, AVX2. For those interested, we suggest to browse the Intel Intrinsics Guide page~\href{https://software.intel.com/sites/landingpage/IntrinsicsGuide}{[link]}.

\subsubsection{Programmable user interfaces: SWIG}
SWIG is a software development tool that connects programs written in C and C++ with a variety of high-level programming languages. SWIG is used with different types of target languages including common scripting languages such as Javascript, Perl, PHP, Python, Tcl and Ruby. The list of supported languages also includes non-scripting languages such as C\#, D, Go language, Java including Android, Lua, OCaml, Octave, Scilab and R. Also several interpreted and compiled Scheme implementations (Guile, MzScheme/Racket) are supported. SWIG is most commonly used to create high-level interpreted or compiled programming environments, user interfaces, and as a tool for testing and prototyping C/C++ software. SWIG is typically used to parse C/C++ interfaces and generate the ``glue code'' required for the above target languages to call into the C/C++ code. SWIG can also export its parse tree in the form of XML. SWIG is free software and the code that SWIG generates is compatible with both commercial and non-commercial projects.

\section{Helper tools}
Besides, Octave, Python, and Maxima, the Linux/Unix environments offer a number of helper tools that can greatly help a scientist in his/her daily job. Here we list a few of them, but we are open to suggestions if your favorite tool is not in this list.
\subsection{Shell commands}
These tools are all ``command-line'' based, that is, they can simply be run in the command line terminal.

\subsubsection{units - conversion program}
Units is a great tool: it knows the value of the most important scientific constants, it performs units conversions, and most of all it's a calculator with units. Let's consider an example: let's computer the average power of a 50 Hz, 300 pC single-bunch charge, 15-GeV beam. One inputs in units the following quantites:
\begin{lstlisting}
$ units -v
You have: 300 pC * 15 GV * 50 Hz
You want: W
\end{lstlisting}
and units returns
\begin{lstlisting}
    300 pC * 15 GV * 50 Hz = 225 W
    300 pC * 15 GV * 50 Hz = (1 / 0.004444444444444444) W
\end{lstlisting}
The option \verb|-v| makes the output of \verb|units| more verbose and more clear.

Units it's an excellent tool that is too often undervalued. The consistent use of it strengthen and simplifies the writing of any physics-based code. Let's compute for example the electric force experienced by two charged particles at a distance, for example. The force can easily be written as:
$$F = K \frac{Q1 \cdot Q2}{d^2}\text{~[eV/m]}$$
where $Q1$ and $Q1$ are obviously the charges of the two particles involved, and $d$ is their relative distance; $K=\frac{1}{4\pi\epsilon_0}$ is the coupling constant.
We choose to use eV/m as units of the force, because expressing the force, e.g. in Newton, would certainly lead to very small numbers. As a general rule, a  good choice is to pick units that make the quantities at play be small numbers whose integer part is larger than 1. We use $e$ (the charge of a positron) as the units of charge, and mm as the units of distance. Units helps us compute the numerical value of the coupling constant $K$, in the desired units:
\begin{lstlisting}
$ units -v
You have: e*e / 4 pi epsilon0 mm^2
You want: eV/m
	e*e / 4 pi epsilon0 mm^2 = 0.001439964547846902 eV/m
	e*e / 4 pi epsilon0 mm^2 = (1 / 694.4615417756247) eV/m
\end{lstlisting}
Therefore our code will look like:
\begin{lstlisting}
Q1 = -1; % the charge of an electron [e]
Q2 = +1; % the charge of a proton [e]
d = 1; % the relative distance [mm]
K = 0.001439964547846902; % the coupling constanct [e*e / 4 pi epsilon0 mm^2]
\end{lstlisting}
The result being:
\begin{lstlisting}
F = K * Q1 * Q2 / (d*d) % the force [eV/m]
\end{lstlisting}

\subsubsection{bc - An arbitrary precision calculator language}
\verb|bc| is a shell calculator that supports arbitrary precision numbers with interactive execution of statements. There are some similarities in the syntax to the C programming language. A standard math library is available by command line option. If requested, the math library is defined before processing any files. 

In \verb|bc|, the variable \verb|scale| allows one to select the total number of decimal digits after the decimal point to be used:
\begin{lstlisting}
$ bc 
bc 1.06
Copyright 1991-1994, 1997, 1998, 2000 Free Software Foundation, Inc.
This is free software with ABSOLUTELY NO WARRANTY.
For details type `warranty'.
scale=1
sqrt(2)
1.4
scale=40
sqrt(2)
1.4142135623730950488016887242096980785696
\end{lstlisting}

\section{Suggested literature}
There are a number of classic books that every scientist dealing with numerical calculations should know. Here we list some of our favourites:
\begin{itemize}
    \item Donald Knuth, ``\textbf{The Art of Computing programming}'', is a comprehensive monograph written by computer scientist Donald Knuth that covers many kinds of programming algorithms and their analysis. Knuth began the project, originally conceived as a single book with twelve chapters, in 1962. The four volumes are:
    \begin{itemize}
        \item[] Volume 1 – Fundamental Algorithms: \emph{Basic concepts}, \emph{Information structures}
        \item[] Volume 2 - Seminumerical Algorithms: \emph{Random numbers}, \emph{Arithmetic}
        \item[] Volume 3 – Sorting and searching: \emph{Sorting}, \emph{Searching}
        \item[] Volume 4 - Combinatorial searching: \emph{Combinatoiral searching}.
    \end{itemize}
    \textit{En passant}, Donald Knuth is the creator of \TeX, the typesetting system at the base of \LaTeX.
    
    \item W. Press, S. Teukolsky, W. Vetterling, and B. Flannery, ``\textbf{Numerical Recipes: The Art of Scientific Computing}'',  is a complete text and reference book on scientific computing. In a self-contained manner it proceeds from mathematical and theoretical considerations to actual practical computer routines. Even though its routines are nowadays available in libraries such as GSL or NAG, this book remains the most practical, comprehensive handbook of scientific computing available today. Cambridge University Press.
    
    \item Abramowitz and Stegun, ``\textbf{Handbook of Mathematical Functions with Formulas}''. Since it was first published in 1964, the 1046-page Handbook has been one of the most comprehensive sources of information on special functions, containing definitions, identities, approximations, plots, and tables of values of numerous functions used in virtually all fields of applied mathematics.
    
    At the time of its publication, the Handbook was an essential resource for practitioners. Nowadays, computer algebra systems have replaced the function tables, but the Handbook remains an important reference source for finite difference methods, numerical integration, etc.
    
    A high quality scan of the book is available at the University of Birmingham, UK \href{https://www.cs.bham.ac.uk/~aps/research/projects/as/}{[link]}.

    \item Olver, F. , Lozier, D. , Boisvert, R. and Clark, C. (2010), ``\textbf{The NIST Handbook of Mathematical Functions}''. This is a modern version of the Abramowitz-Stegun, and is comprehensive collection of mathematical functions, from elementary trigonometric functions to the multitude of special functions.

    \item Zyla,  P.  A.,  \emph{et al.}, ``\textbf{Review  of  Particle  Physics}'', Oxford  University  Press. s. A huse summary of particle physics, enriched with extremely useful reviews of topics such as particle-matter interaction, probability, Monte Carlo techniques, and statistics.

    \item George B. Arfken, ``\textbf{Mathematical Methods for Physicists}''. This is a thorough handbook about mathematics that is useful in physics. It is a venerable book that goes back to 1966; this seventh edition (2012) adds a new co-author, Frank E. Harris. It is pitched as a textbook but because of its size and breadth probably works better as a reference. Very Good Feature: all the examples are real problems from physics. Note that it is not a mathematical physics book, though: it quotes the models but does not develop them.

    \item Hockney, ``\textbf{Computer Simulation Using Particles}''. This is another venerable reference in scientific research, including the simulation of systems through the motion and the interaction of particles. This book provides an introduction to simulation using particles based on the PIC (Particle-in-cell), CIC (Cloud-in-cell), and other algorithms and the programming principles that assist with the preparations of large simulation programs.

\end{itemize}

\end{document}